\documentclass[aip]{revtex4-1}
\usepackage{graphicx}
\usepackage{dcolumn}
\usepackage{bm}
\usepackage{amsmath,amssymb,amsfonts,amsthm,color}
\usepackage{algorithm}
\usepackage{algpseudocode}
\usepackage{subcaption}

\newtheorem{rem}{Remark}

\def\mb{\boldsymbol}
\def\half{\frac{1}{2}}

\def\bm{\boldsymbol}
\def\bma{\boldsymbol}

\def\eg{{\it e.g.}}

\begin{document}
\title{On Consistent Definitions of Momentum and Energy Fluxes for Molecular Dynamics Models with  Multi-body Interatomic Potentials}
\author{Xiaojie Wu}
\email{xxw139@psu.edu}
\affiliation{Department of Mathematics, The Pennsylvania State University.}
\author{Xiantao Li}
\email{xli@math.psu.edu}
\affiliation{Department of Mathematics, The Pennsylvania State University.}

\begin{abstract}
 Results from molecular dynamics simulations often need to be further processed to understand the physics at a larger scale. This paper considers the definitions of momentum and energy fluxes obtained from a control-volume approach.  To assess the validity of these defined quantities, two consistency criteria are proposed. As examples, the embedded atom potential and the Tersoff potential are considered. The consistency is verified by analytical and numerical methods.
\end{abstract}

\maketitle

\section{Introduction}
Molecular dynamics (MD) plays a unique role in the modeling and simulation of modern material science problems \cite{FrSm02,rapaport2004art}.
In principle, it is convenient at this level to introduce lattice structures and material defects, and then study their implications to the overall mechanical and thermal properties. Examples include crack propagation, dislocation dynamics, energy conduction, etc. An important step in MD-based simulations is the calculation of quantities of interest based on particle trajectories. In particular, it provides  the connection between molecular trajectories and processes on the macro- and mesoscopic scales.

The main purpose of this paper is to discuss  consistent definitions of momentum and energy fluxes, both of which are essential components of continuum thermoelasticity models \cite{nowacki1975dynamic} and micro-polar models \cite{eringen1999theory}. The computation of the mechanical quantities from molecular-level description is by no means a new concept. Perhaps the earliest works date back to  Clausius and Maxwell \cite{clausius1870xvi, maxwell1870reciprocal, clerk1874ver} in 1870. Thanks to the recent papers \cite{Tsai1979virial, Lutsko1988stress, Zhou2002equivalent, Zhou2003new, Zhou2005thermomechanical, Cormier2001stress, Cheung1991atomic},  many interesting issues have been brought to light. This current paper does not attempt to review these important contributions. 
Rather, we choose to address the issue of consistency to a greater extent.
More specifically, we think about the consistency at two levels. First, they should be consistent with fundamental conservation laws. Such consistency can be ensured by following the Irving-Kirkwood approach\cite{Irving1950}, a particular example of which is the Hardy's derivation \cite{Hardy1982formulas, Hardy2003shock}. Hardy's approach has been implemented and improved in various different ways by several groups \cite{Chen2006local, Yang2012generalized, Yang2014accurate, Zimmerman2004calculation, Zimmerman2010material,  Fu2013modification, Fu2013evaluation}. In this paper, however, we follow the control-volume approach \cite{Smith2012control}, which typically is the starting point for deriving a continuum mechanics model.  As a result, we obtain expressions for the momentum flux --- the traction,  and the energy flux for the interfaces between the control volumes, which can be viewed as a finite-volume representation of the fundamental conservation laws.

 Meanwhile, it is well known that even when the fundamental conservation laws are obeyed, there are still ambiguities in defining  these continuum quantities \cite{Admal2011stress}. More specifically, the expressions depend on how the interatomic forces and energy are partitioned among the atoms. This is clearly an alarming issue, and  has motivated us to impose a second level of consistency:  i.e., the consistency with the continuum limit. For crystalline solids, the continuum limit is in the form of the elastic wave equations, augmented with the Cauchy-Born rule\cite{CB84,ArGr05,Weinan2007cauchy}.  The Cauchy-Born rule in principle does not depend on how the force and energy are decomposed. As a result, it provides an alternative guideline for the consistency check.  In this paper, we formulate this  criterion and examine the consistency both analytically and numerically.

For MD models with pairwise interaction, the calculation of elastic stress, or traction, the projection of the stress to a specific interface, is relatively easy. For multi-body interactions, however, the issue is much more complicated \cite{Chen2006local, Admal2011stress,Admal2010unified}. On one hand, the formulas that are directly generalized from pair potentials may not satisfy the consistency criteria  postulated here. On the other hand, even though several formulas have been derived  based on conservation laws \cite{Chen2006local, Admal2011stress,Admal2010unified}, the second consistency criterion has not been evaluated.  


To address the issues of consistency with sufficient specificity, we consider two concrete examples: The embedded atoms model (EAM) \cite{Daw1984eam} and the Tersoff potential\cite{Tersoff1988new}, which are among the most popular empirical potentials in modern molecular simulations. For each of these two models, we discuss how to compute the traction and energy flux within the molecular simulation. The  consistency at both levels is carefully assessed.  Furthermore, for the Tersoff potential, we provide a  pseudo code for the calculation of the traction and energy flux to help interested readers to implement the formulas.

The paper is organized as follows. First, we introduce the general framework for the control-volume approach, and demonstrate how the traction and energy flux arise under this framework. Then, we focus on the explicit expressions for the traction and energy flux for the EAM and Tersoff potentials. The consistency criteria are discussed in sections \ref{sec: tests} and \ref{sec: consis2}. In the appendix, we provide the pseudo code for the Tersoff potential.

\section{The Derivation of The Traction and Energy Flux}

Our definition of all the quantities is based on the coordinates and velocity of the atoms, here denoted by $\bm x_i(t)$ and $\bm v_i(t)(=\dot{\bm x}_i(t))$. The trajectory of the atoms is determined from the molecular dynamics model (MD), 
\begin{equation}\label{eq: md}
 m_i \ddot{\bm x}_i = \bm f_i, \quad \bm f_i=-\frac{\partial V}{\partial \bm x_i}.
\end{equation}
Here $V$ is the interatomic potential, which we will assume to be an empirical model.

\begin{figure}[htbp]
\begin{center}
\includegraphics[scale=0.54]{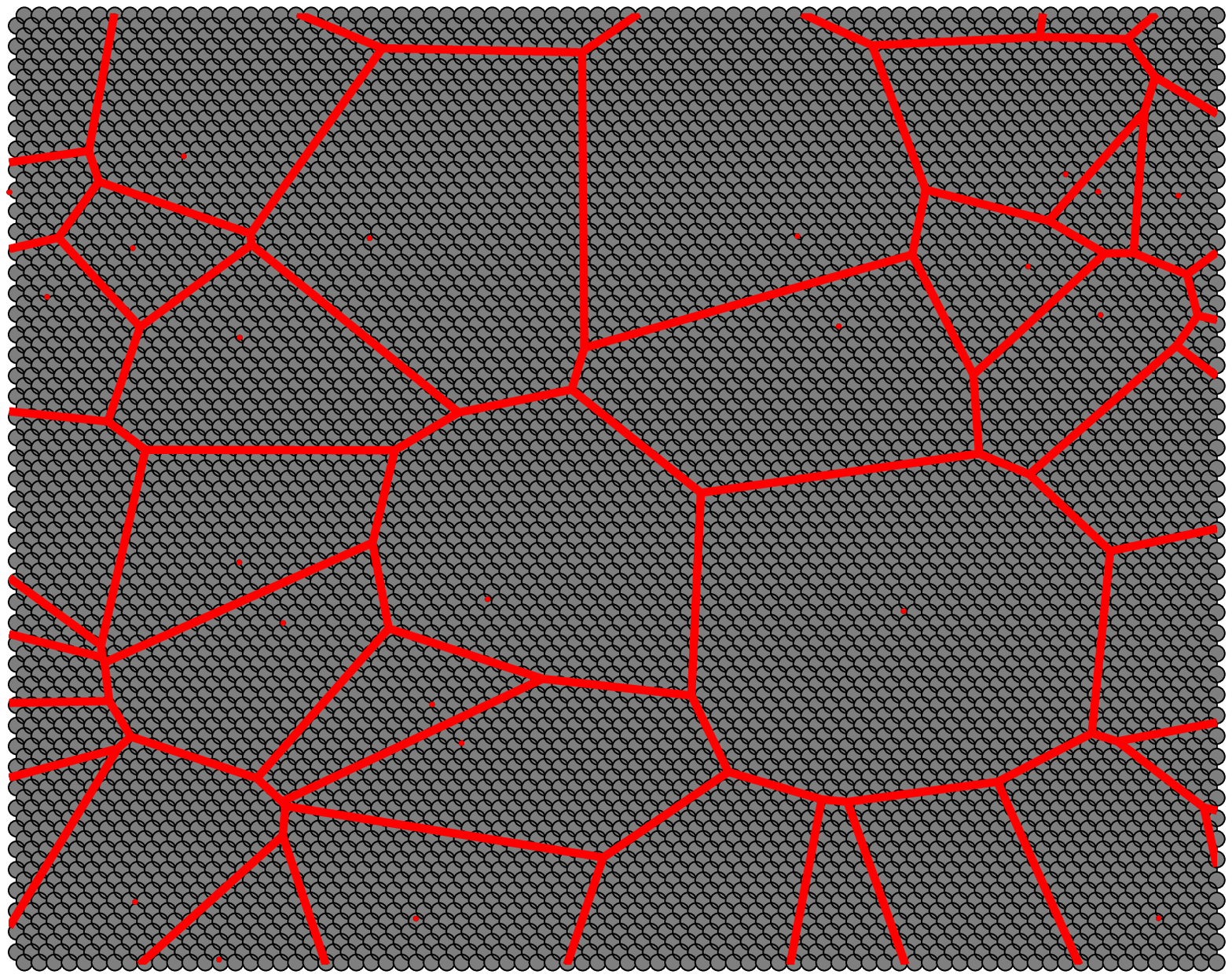}
\caption{The partition of the system: The atoms are grouped into different cells.}
\label{fig: div}
\end{center}
\end{figure}

Our approach has been motivated by the mathematical formulation
of continuum mechanics models, which usually starts with a control volume and then derives
the equations based on mass, momentum, and energy balance. To follow this approach, we divide the system into separate cells, each denoted by $\Omega_\alpha$, as illustrated in Figure \ref{fig: div}. 

The key observation is that the rate of change of the moment and energy in each cell is determined
by the momentum and energy fluxes across the cell interfaces. Let us first consider the momentum balance. Toward this end, we define the total momentum  in the cell $\Omega_\alpha$,
\begin{equation}\label{eq: tot-momen}
\bm p_\alpha(t)= \sum_{i \in \Omega_\alpha} m_i \bm v_i(t). 
\end{equation}
Here we follow the reference (Lagrangian) coordinate. Namely, the notation $i \in \Omega_\alpha$ indicates that the reference position of the $i$th atom is in the cell $\Omega_\alpha$.
Our choice is the same as the material frame used in\cite{Zimmerman2010material}, while most other derivations are based on current (Eulerian) coordinates.

Similarly,  we define a local energy,
\begin{equation}\label{eq: E-alpha}
 E_\alpha(t)= V_\alpha + \sum_{i \in \Omega_\alpha} \frac{\bm p_i^2}{2m_i},
\end{equation}
where $V_\alpha$ is the potential energy in the domain $\Omega_\alpha$, whose definition will later be 
made more precise.

With the local momentum and energy selected based on the position and velocity of the atoms, we  now seek to define a momentum flux (traction) and energy flux between the cell $\Omega_\alpha$ and
a neighboring cell $\Omega_{\beta}$. These fluxes, denoted by $\bm t_{\alpha,\beta}$ and 
$q_{\alpha,\beta}$, respectively, are required to satisfy the following four conditions. 
\begin{enumerate}
\item [(i)] {\bf Compatibility: } The definitions should follow from the MD model \eqref{eq: md}. As a result, the expressions should depend on the functional forms of $V$, and the involved parameters;
\item [(ii)] {\bf Conservativeness:} $\bm t_{\alpha, \beta}= - \bm t_{\beta, \alpha}$ and $q_{\alpha, \beta}= - q_{\beta, \alpha}$, as motivated by the finite-volume methods for conservation laws \cite{LeVeque03};
\item [(iii)] {\bf Momentum and energy balance:}
  \[ 
  \begin{aligned}
   \frac{d}{dt}\bm p_\alpha(t)=& \sum_{\beta} \bm t_{\alpha, \beta},\\
   \frac{d}{dt}E_\alpha(t)=& \sum_{\beta} q_{\alpha, \beta};
   \end{aligned}
   \]
\item [(iv)] {\bf Locality:} For an empirical potential with short-range interactions,  $\bm t_{\alpha, \beta}$ and  $q_{\alpha, \beta}$ should only depend on the atoms near the interface between $\Omega_\alpha$ and $\Omega_{\beta}$.   
\end{enumerate}

 In general, for the momentum balance, the key ingredient is a decomposition of the force,
\begin{equation}\label{eq: fij}
  \bm f_i = \sum_{j} \bm f_{ij}, 
\end{equation}
with the property that \(\bm f_{ij}= -\bm f_{ji}\). This, however, does not imply that the interatomic
interaction is pairwise. In fact, the force $\bm f_{ij}$ may depend on other atoms. 

Combining \eqref{eq: tot-momen} and \eqref{eq: fij}, we find that,
\[ \frac{d}{dt} \bm p_\alpha(t)= \sum_{i \in \Omega_\alpha} \sum_j \bm f_{ij}.\] 
Notice that due to the asymmetry of $\bm f_{ij}$, we have
\[ \sum_{i \in \Omega_\alpha} \sum_{j \in \Omega_\alpha} \bm f_{ij}=0.\]
As a result, we can restrict $j$ to the outside of $\Omega_\alpha$,
\begin{equation}
 \frac{d}{dt} \bm p_\alpha(t)= \sum_{i \in \Omega_\alpha}\sum_{j \notin \Omega_\alpha} \bm f_{ij},
\end{equation}
which leads naturally to a definition of the traction,
\begin{equation}
 \bm t_{\alpha,\beta}= \sum_{i \in \Omega_\alpha}\sum_{j \in \Omega_\beta} \bm f_{ij}.
\end{equation}
This traction defined this way  clearly satisfies the conditions (i), (ii), and (iii). The forth condition needs to be
checked for a specific model. Another important issue, as raised by Admal and Tadmor\cite{Admal2010unified}, is that the force decomposition is generally not unique. We will defer this discussion to the section \ref{sec: consis2}, where we discuss the second-level consistency.

\smallskip

Meanwhile, for the energy balance, the difficulty is to divide the potential energy among the atoms.
We need to define energy $V_i$, such that 
\begin{equation}\label{eq: vsum}
V= \sum_i V_i.
\end{equation}
 Once we have this energy partition at hand,
we define
\begin{equation}
 V_\alpha= \sum_{i \in \Omega_\alpha} V_i.
\end{equation}

In the next two subsections, we will discuss the derivations for two specific empirical potentials.

\subsection{The Embedded Atom Potential}

The first model to be considered is the embedded atom model (EAM) \cite{Daw1984eam}:
\begin{equation}\label{eq: eam}
 V_\text{EAM} = \half \sum_{1\le i \le N} \sum_{1\le j \le N, j\ne i} \varphi(r_{ij}) 
 + \sum_{1\le i \le N} E(\rho_i), \quad \rho_i= \sum_{1\le j \le N, j\ne i} \rho(r_{ij}).  
\end{equation}
Here we have adopted the usual notation in molecular simulations: $\bm r_{ij}= \bm r_i - \bm r_j$, and $r_{ij}= |\bm r_{ij}|.$ The function $\rho$ represents the influence of local electron density, and because of the nonlinearity of the function $E$, the interaction is of multi-body nature.

\subsubsection{The Definition of the Traction}

For the EAM potential \eqref{eq: eam}, the force on the atom $i$ can be obtained with direct calculations, and it is given by,
\begin{equation}\label{eq: eamf}
 \bm f_i = \sum_{j\ne i} -\Big\{ \varphi'(r_{ij}) + \rho'(r_{ij})\big[E'(\rho_i)+ E'(\rho_j)\big]\Big\}\frac{\bm r_{ij}}{r_{ij}}.
\end{equation}
The most commonly used  (and perhaps the most natural ) force decomposition is as follows,
\begin{equation}\label{eq: fu}
 \bm f_{ij} = -\Big\{\varphi'(r_{ij}) + \rho'(r_{ij})\big[E'(\rho_i)+ E'(\rho_j)\big]\Big\}\frac{\bm r_{ij}}{r_{ij}}.
\end{equation}
It clearly satisfies the two conditions in \eqref{eq: fij}. This leads to the definition of the traction,
\begin{equation}\label{eq: t-eam}
\bm t_{\alpha,\beta}= \sum_{i \in \Omega_\alpha} \sum_{j \in \Omega_\beta}  -\Big\{\varphi'(r_{ij}) + \rho'(r_{ij})\big[E'(\rho_i)+ E'(\rho_j)\big]\Big\}\frac{\bm r_{ij}}{r_{ij}}.
\end{equation}

\subsubsection{The Definition of the Energy Flux}

To obtain the energy flux, we first choose,
\begin{equation}
V_i= \half \sum_{j\ne i} \varphi(r_{ij}) + E(\rho_i).
\end{equation}
For the pair interaction $\varphi(r_{ij})$, we split the energy among the two atoms equally.
The second part comes from the embedded energy, and these energy terms clearly add up to the total potential energy $V$, i.e., equation \eqref{eq: vsum} is satisfied.

To continue, we start with \eqref{eq: E-alpha}, and calculate the time derivatives of the
kinetic and potential energy as follows:
\begin{align*}
\frac{d}{dt}\sum_{i\in \Omega_\alpha}\frac{1}{2}m_i \bm v_i^2 
&=\sum_{i\in \Omega_\alpha}\bm f_i\cdot\bm v_i\\
&=-\sum_{i\in \Omega_\alpha}\sum_{j\neq i}\big[E'(\rho_i)+E'(\rho_j)\big]\rho'(r_{ij})\frac{\bm r_{ij}\cdot \bm v_i}{r_{ij}}-\sum_{i \in \Omega_\alpha}\sum_{j\neq i}\varphi'(r_{ij})\frac{\bm r_{ij}\cdot \bm v_i}{r_{ij}},
\end{align*}
and,
\begin{align*}
\frac{d}{dt}\sum_{i\in \Omega_\alpha}V_i=&\sum_{i\in \Omega_\alpha}E'(\rho_i)\sum_{j\neq i}\rho'(r_{ij})\frac{\bm r_{ij}\cdot(\bm v_i-\bm v_j)}{r_{ij}}+\frac{1}{2}\sum_{j\neq i}\varphi'(r_{ij})\frac{\bm r_{ij}\cdot(\bm v_i-\bm v_j)}{r_{ij}}\\
=&\sum_{i\in \Omega_\alpha}E'(\rho_i)\sum_{j\neq i}\rho'(r_{ij})\frac{\bm r_{ij}\cdot\bm v_i}{r_{ij}} - \sum_{i\in \Omega_\alpha}E'(\rho_i)\sum_{j\neq i}\rho'(r_{ij})\frac{\bm r_{ij}\cdot\bm v_j}{r_{ij}}\\
&+\frac{1}{2}\sum_{i\in \Omega_\alpha}\sum_{j\neq i}\varphi'(r_{ij})\frac{\bm r_{ij}\cdot\bm v_i}{r_{ij}} - \frac{1}{2}\sum_{i\in \Omega_\alpha}\sum_{j\neq i}\varphi'(r_{ij})\frac{\bm r_{ij}\cdot\bm v_j}{r_{ij}}.
\end{align*}

Combining terms, we get 
\begin{align*}
&\frac{d}{dt}\sum_{i\in \Omega_\alpha}(E_i+\frac{1}{2}m_i \bm v_i^2) \\
= & -\frac{1}{2}\sum_{i \in \Omega_\alpha}\sum_{j \neq i}\phi'(r_{ij})\frac{\bm r_{ij}\cdot(\bm v_i+\bm v_j)}{r_{ij}} \\ - &
\sum_{i \in \Omega_\alpha} \sum_{j\neq i}\Big\{ E'(\rho_j) \rho'(r_{ij})\frac{\bm r_{ij}\cdot\bm v_i}{r_{ij}} + E'(\rho_i)\rho'(r_{ij})\frac{\bm r_{ij}\cdot\bm v_j}{r_{ij}}\Big\}.
\end{align*}

Now we can easily see that within the summation terms on the right hand side, each term changes sign 
when the indices \(i\) and \(j\) are exchanged. Hence, we can restrict \(j\) to outside \(\Omega_\alpha\), and  the energy flux can be defined as,
\begin{equation}\label{eq: qeam}
 \begin{aligned}
 q_{\alpha,\beta}= &\half\sum_{i\in \Omega_{\alpha}, j\in \Omega_\beta} \varphi'(r_{ij})\frac{\bm r_{ij}}{r_{ij}} \cdot (\bm v_i + \bm v_j) \\
 &- \sum_{i\in \Omega_\alpha, j\in \Omega_\beta}\big[(E'(\rho_i) \bm v_j + E'(\rho_j) \bm v_i\big]\cdot \frac{\bm r_{ij}}{r_{ij}}.
\end{aligned}
\end{equation}

It can be directly verified that this energy flux satisfies all the requirement listed in the previous section. 

\begin{rem} In the case when $E\equiv 0$, i.e., the EAM potential is reduced to a pair potential, the energy flux can be written in a compact form,
\begin{equation}
 q_{\alpha,\beta}= \half\sum_{i\in \Omega_{\alpha}, j\in \Omega_\beta} \bm f_{ij} \cdot (\bm v_i + \bm v_j).
\end{equation}
Unfortunately, for multi-body interactions, this formula is no longer correct and should not be used in practice. This can be seen from  \eqref{eq: qeam}.
\end{rem}

\bigskip

\subsection{The Tersoff Potential}
Another important empirical potential is the Tersoff potential \cite{Tersoff1988new, Tersoff1989modeling} consisting of a pairwise interaction and a multi-body interaction,
\begin{equation}
E= \sum_{i\ne j} \Big[ \frac12 f_{RC}(r_{ij}) + V_{ij}\Big]
\end{equation}
Since the pair potential $f_{RC}(r)$ is much easier to work with, we will only focus on the second
part. 

This model may seem to be a three-body interaction, but the interaction is actually among more neighboring atoms. Therefore, the total energy can not be written as,
\[ V= \sum_{i,j,k} V(\bm r_i, \bm r_j, \bm r_k).\] 
 Due to the complexity of the function forms, we will derive the formulas in several steps. The calculation is a bit lengthy, but we choose to show all the steps for the purpose of mathematical clarity. We are not aware of any simpler derivations. 

\smallskip

Following \cite{Tersoff1988new, Tersoff1989modeling},  we express the multi-body term as follows,
\begin{equation}
V_{ij}=\frac12 f_{AC}(r_{ij}) B(\zeta_{ij}), \quad \zeta_{ij}= \sum_{k\ne i,j} V_3(\bma r_{ij}, \bma r_{ik}),  
\end{equation}
where we have defined,
\begin{equation}
 V_3(\bma u, \bma v) =  f_C(v) g\big( c(\bma u, \bma v)\big), \quad c(\bma u, \bma v)= \frac{\bma u \cdot \bma v}{uv}.
\end{equation}

We can compute the interatomic forces due to $V_{ij}$ as follows,
\begin{equation}\label{eq: dvij}
\begin{aligned}
 \frac{\partial V_{ij}}{\partial \bma r_i}=& \frac12 f_{AC}'(r_{ij}) B(\zeta_{ij})\frac{\bma r_{ij}}{r_{ij}} \\
 &+ \frac12 f_{AC}(r_{ij}) B'(\zeta_{ij}) \sum_{k\ne i, j}\Big[ \frac{\partial V_3(\bma r_{ij}, \bma r_{ik})}{\partial \bma r_{ij}}
                                           +  \frac{\partial V_3(\bma r_{ij}, \bma r_{ik})}{\partial \bma r_{ik}}\Big],\\
 \frac{\partial V_{ij}}{\partial \bma r_j}=&  -\frac12 f_{AC}'(r_{ij}) B(\zeta_{ij})\frac{\bma r_{ij}}{r_{ij}} 
 -\frac12 f_{AC}(r_{ij}) B'(\zeta_{ij}) \sum_{k\ne i, j} \frac{\partial V_3(\bma r_{ij}, \bma r_{ik})}{\partial \bma r_{ij}},\\        
 \frac{\partial V_{ij}}{\partial \bma r_k}=&  -\frac12 f_{AC}(r_{ij}) B'(\zeta_{ij}) 
 \frac{\partial V_3(\bma r_{ij}, \bma r_{ik})}{\partial \bma r_{ik}}, \quad \text{for any} \;k\ne i, j.
\end{aligned}
\end{equation}

One can easily check that,
\begin{equation}\label{eq: uv}
\begin{aligned}
 \frac{\partial V_3(\bma u, \bma v)}{\partial \bma u}&= -\frac{f_C(v) g'(c)c}{u^2}\bma u + \frac{f_C(v) g'(c) }{uv} \bma v\\& \stackrel{\rm{def}}{=}  s_{11}(\bma u, \bma v) \bma u
 + s_{12}(\bma u, \bma v) \bma v,\\
 \frac{\partial V_3(\bma u, \bma v)}{\partial \bma v}&=\frac{f_C(v)g'(c)}{uv}\bma u + \Big[ \frac{f_C'(v)g(c)}{v} - \frac{f_C(v)g'(c)c}{v^2}\Big]\bma v\\
 &\stackrel{\rm{def}}{=}  s_{21}(\bma u, \bma v) \bma u
 + s_{22}(\bma u, \bma v) \bma v.
\end{aligned}
\end{equation}
Notice that $s_{12}=s_{21}.$

\medskip

To arrive at an appropriate force decomposition of the form \eqref{eq: fij}, we first make the observation that by properly re-organizing terms using \eqref{eq: uv},  the equation \eqref{eq: dvij} can be written as,
\begin{equation}\label{eq: dVij}
\begin{aligned}
 -\frac{\partial V_{ij}}{\partial \bm r_i} =&  \bm f_{ij, ij} + \sum_{k\ne i, k\ne j} \bm f_{ij,ik},\\
 -\frac{\partial V_{ij}}{\partial \bm r_j} =&  \bm f_{ij, ji} + \sum_{k\ne i, k\ne j} \bm f_{ij,jk},\\
 -\frac{\partial V_{ij}}{\partial \bm r_k} =&  \bm f_{ij, ki} + \bm f_{ij,kj},\quad \text{for}\; k \ne i, k\ne j,\\
\end{aligned}
\end{equation}
where,
\begin{equation}
\left\{
\begin{aligned}
  \bm f_{ij, ij} &= -\frac12 f_{AC}'(r_{ij}) B(\zeta_{ij})\frac{\bma r_{ij}}{r_{ij}} 
     - \frac12 f_{AC}(r_{ij}) B'(\zeta_{ij}) \sum_{k\ne i, j} \big(s_{11} + s_{12} \big) \bm r_{ij}, \\
  \bm f_{ij, ik} &=     - \frac12 f_{AC}(r_{ij}) B'(\zeta_{ij}) \big(s_{12} + s_{22} \big) \bm r_{ik}, \\
  \bm f_{ij, ji} &=   -\frac12 f_{AC}'(r_{ij}) B(\zeta_{ij})\frac{\bma r_{ji}}{r_{ji}}     
     - \frac12 f_{AC}(r_{ij}) B'(\zeta_{ij}) \sum_{k\ne i, j} \big(s_{11} + s_{12} \big) \bm r_{ji}, \\
  \bm f_{ij, jk} &=      \frac12 f_{AC}(r_{ij}) B'(\zeta_{ij}) s_{12} \bm r_{jk}, \\
  \bm f_{ij, ki} &= -\frac12 f_{AC}(r_{ij}) B'(\zeta_{ij}) \big(s_{12} + s_{22} \big) \bm r_{ki}, \\
  \bm f_{ij, kj} &=      \frac12 f_{AC}(r_{ij}) B'(\zeta_{ij}) s_{12} \bm r_{kj}. \\
\end{aligned}\right.
\end{equation}
These force components are defined in such a way to ensure that (a) they are anti-symmetry, \eg, $\bm f_{ij,ik}= -\bm f_{ij,ki}$; (b) they are central forces, \eg, $\bm f_{ij,jk} \parallel \bm r_{jk}.$  

As a result, to obtain a force decomposition of the general form \eqref{eq: fij}, the force on atom $i$ is written  as follows,
\begin{equation}\label{eq: fi}
 \begin{aligned}
\bm f_i=& \sum_{j \ne i}\big[\bm f_{ij,ij}+\bm f_{ji,ij}\big] + \sum_{j \ne i} \sum_{k\ne i, k\ne j} \big[\bm f_{ij,ik}+\bm f_{ji,ik}\big] \\
       +&\frac12\sum_{k \ne i, \ell \ne i} \big[\bm f_{k \ell,ik}+\bm f_{\ell k,ik}+\bm f_{k\ell,i\ell}+\bm f_{\ell k,i\ell}\big]. 
  \end{aligned}         
\end{equation}
The first term is viewed as the direct interaction between atoms $i$ and $j$. The second term includes forces pointing to other atoms. The last term contains the forces due to other pairs of atoms that do not involve atom $i$. Notice that 
the first two terms can be combined. But we will keep them separate for make the following calculations more transparent.
In addition, the factor $\half$ takes into account of double counting. It is also written this way to make later calculations easier.

\subsubsection{Definition of the Traction}

We first derive the traction following \eqref{eq: tot-momen}. From \eqref{eq: fi}, we start with the first sum,
\[ \sum_{i \in \Omega_\alpha} \sum_{j \ne i} \big[\bm f_{ij,ij}+\bm f_{ji,ij}\big]
= \sum_{i \in \Omega_\alpha} \sum_{j \notin \Omega_\alpha} \big[\bm f_{ij,ij}+\bm f_{ji,ij}\big].\]
So we define,
\begin{equation}\label{eq: t-I}
\bm t_{\alpha,\beta}^I= \sum_{i \in \Omega_\alpha} \sum_{j \in \Omega_\beta} \big[\bm f_{ij,ij}+\bm f_{ji,ij}\big].
\end{equation}

The remaining terms in the force \eqref{eq: fi} will be combined as follows,
\[ 
\sum_{i \in \Omega_\alpha} \sum_{j\ne i, k \ne i, j\ne k} \Big[ \bm f_{ij, ik} + \bm f_{ji,ik} + \bm f_{jk, ij}
+ \bm f_{jk,ik} \Big].\]
Notice that for the third term in  \eqref{eq: fi}, we changed the index $\ell$ to $j$.

Let us use the fact that the summation over the follow range is zero:
\[ \sum_{i \in \Omega_\alpha}\sum_{j\in \Omega_\alpha} \sum_{k\in \Omega_\alpha} =0.\]
This can be verified by changing the indices \((i,j,k)\) to \((j,k,i)\) for the first term, and  \((i,j,k)\) to \((k,j,i)\)
for the second term. Thus, we can split the triple sum into three:
\[ \sum_{i \in \Omega_\alpha} \sum_{j\ne i, k \ne i, j\ne k} =
\sum_{i \in \Omega_\alpha} \sum_{j\in \Omega_\alpha} \sum_{k\in \Omega_\beta}  
\quad +\sum_{i \in \Omega_\alpha}\sum_{j\in \Omega_\beta} \sum_{k\in \Omega_\alpha} 
\quad +\sum_{i \in \Omega_\alpha}\sum_{j\in \Omega_\beta} \sum_{k\in \Omega_\beta}  .\]

We collect the first two terms in the first sum and the last two terms in the third sum. We define,
\begin{equation}\label{eq: t-II}
 \bm t_{\alpha, \beta}^{II}= 
 \sum_{i \in \Omega_\alpha} \sum_{j\in \Omega_\alpha} \sum_{k\in \Omega_\beta} \big[
  \bm f_{ij, ik} + \bm f_{ji,ik} \big]
-   \sum_{i \in \Omega_\alpha}\sum_{j\in \Omega_\beta} \sum_{k\in \Omega_\beta}\big[
\bm f_{jk, ij}
+ \bm f_{jk,ik} \big].
\end{equation}

Meanwhile, the second sum can be simplified to, 
\[ \sum_{i \in \Omega_\alpha}\sum_{j\in \Omega_\beta} \sum_{k\in \Omega_\alpha}
\Big[ \bm f_{ij, ik} +  \bm f_{jk, ij} \Big].\]
The second and forth terms cancel if we change the indices  \((i,j,k)\) to \((k,j,i)\) in the second term. 
Therefore, all the remaining terms can be written as,
\[
\begin{aligned}
 \bm t_{\alpha, \beta}^{III}= & \sum_{i \in \Omega_\alpha} \sum_{j\in \Omega_\alpha} \sum_{k\in \Omega_\beta}   
 \big[\bm f_{jk, ij}
+ \bm f_{jk,ik} \big] \\
 &+ \sum_{i \in \Omega_\alpha}\sum_{j\in \Omega_\beta} \sum_{k\in \Omega_\alpha}
\Big[ \bm f_{ij, ik} +  \bm f_{jk, ij} \Big] \\
 & +\sum_{i \in \Omega_\alpha}\sum_{j\in \Omega_\beta} \sum_{k\in \Omega_\beta} 
\big[ \bm f_{ij, ik} + \bm f_{ji,ik} \big].
\end{aligned}
\]
We further notice that the first term in the first sum and the first term in the second sum cancel. This becomes apparent 
if the indices for the second term are changed from \((i,j,k)\) to \((j,k,i)\).
As a result, we may simplify this part of the traction:
\begin{equation}\label{eq: t-III}
 \bm t_{\alpha, \beta}^{III}= \sum_{i \in \Omega_\alpha}\sum_{j\in \Omega_\beta} \sum_{k\in \Omega_\beta}  \big[
 \bm f_{ij,ik} + \bm f_{ji,ik}\big] 
 - \sum_{i \in \Omega_\alpha}\sum_{j\in \Omega_\beta} \sum_{k\in \Omega_\alpha} 
 \big[ \bm f_{ij,jk} + \bm f_{ji,jk}\big].
\end{equation}
The last two terms are obtained from the equation above, by changing indices \((j,k,i)\) to \((i,j,k)\) for the second term in the first sum, and exchanging the indices $i$ and $k$  for the second term in the second sum. We can verify that the traction defined this way satisfy all the conditions listed
in the previous section.

\subsubsection{Definition of the Energy Flux}

Next, we define a local energy as follows,
\begin{equation}\label{eq: V-i}
 V_i= \frac{1}{2}\sum_{  j\ne i} \big[V_{ij} + V_{ji}\big].
\end{equation}

The energy flux will be reflected in the rate of energy change. In particular, we will compute $\frac{d}{dt}E_\alpha(t).$ It
is clear that the derivative of the kinetic energy is as follows,
\begin{equation}\label{eq: kin}
\begin{aligned}
 \frac{d}{dt} \sum_{i \in \Omega_\alpha} \frac{\bm p_i^2}{2m_i} =& \sum_{i \in \Omega_\alpha} \bm f_i \cdot \bm v_i \\
 =& \sum_{j \ne i}\big[\bm f_{ij,ij}+\bm f_{ji,ij}\big]\cdot \bm v_i + \sum_{j \ne i} \sum_{k\ne i, k\ne j} \big[\bm f_{ij,ik}+\bm f_{ji,ik}\big]\cdot \bm v_i \\
       +&\frac12\sum_{k \ne i, \ell \ne i} \big[\bm f_{k \ell,ik}+\bm f_{\ell k,ik}+\bm f_{k\ell,i\ell}+\bm f_{\ell k,i\ell}\big]\cdot \bm v_i. 
  \end{aligned}   
\end{equation}

To calculate the change of the potential energy, we begin with,
\begin{equation}\label{eq: pot}
 \begin{aligned}
\frac12\frac{d}{dt}\big(V_{ij}+V_{ji}\big)=& -\half\big[ \bm f_{ij,ij} +  \bm f_{ji,ij}\big] \cdot \bm v_i -\half \sum_{k\ne i, k\ne j} \big[ \bm f_{ij,ik} +  \bm f_{ji,ik}\big] \cdot \bm v_i \\
 & -\half\big[ \bm f_{ij,ji} +  \bm f_{ji,ji}\big] \cdot \bm v_j - \half\sum_{k\ne j,k\ne i} \big[\bm f_{ij,jk} +  \bm f_{ji,jk} \big]\cdot \bm v_j \\
   &- \half\sum_{k\ne i, k\ne j}  \big[\bm f_{ij,ki} + \bm f_{ji,ki} + \bm f_{ij, kj}+\bm f_{ji, kj}\big]\cdot \bm v_k.
  \end{aligned} 
\end{equation}

We now combine \eqref{eq: pot} and \eqref{eq: kin}. We start by collecting similar terms. We first have,
\[
\begin{aligned}
&\sum_{i \in \Omega_\alpha} \sum_{j \ne i}\big[\bm f_{ij,ij}+\bm f_{ji,ij}\big]\cdot \bm v_i
 -\half\big[ \bm f_{ij,ij} +  \bm f_{ji,ij}\big] \cdot \bm v_i 
 - \half\big[ \bm f_{ij,ji} +  \bm f_{ji,ji}\big] \cdot \bm v_j \\
&= \sum_{i \in \Omega_\alpha} \sum_{j \ne i} \half \Big\{ \big[\bm f_{ij,ij}+\bm f_{ji,ij}\big]
\cdot \bm v_i - \big[\bm f_{ij,ji}+\bm f_{ji,ji}\big]
\cdot \bm v_j\Big\}\\
&= \sum_{i \in \Omega_\alpha} \sum_{j \notin \Omega_\alpha}\half \Big\{ \big[\bm f_{ij,ij}+\bm f_{ji,ij}\big]
\cdot \bm v_i - \big[\bm f_{ij,ji}+\bm f_{ji,ji}\big]
\cdot \bm v_j\Big\}.   
\end{aligned}
\] 
This expression can be further simplified to, 
\begin{equation}\sum_{i \in \Omega_\alpha} \sum_{j \notin \Omega_\alpha}\half \big[\bm f_{ij,ij}+\bm f_{ji,ij}\big]
\cdot \big[\bm v_i + \bm v_j\big]. 
\end{equation}
But this formula only holds for this part of the flux.

Here we have used the fact that when $j \in \Omega_\alpha$, the terms in the bracket ($\{\cdot\}$) would cancel. As a result,
this flux is only dependent on atoms near the boundary of $\Omega_\alpha$. We define for two neighboring cells
$\Omega_\alpha$ and $\Omega_\beta$,
\begin{equation}\label{eq: q-I}
\bm q_{\alpha,\beta}^I=\half \sum_{i \in \Omega_\alpha} \sum_{j \in \Omega_\beta}\Big\{ \big[\bm f_{ij,ij}+\bm f_{ji,ij}\big]
\cdot \bm v_i - \big[\bm f_{ij,ji}+\bm f_{ji,ji}\big]
\cdot \bm v_j\Big\}. 
\end{equation}
\smallskip

We proceed with the following terms,
\[
\begin{aligned}
&\sum_{i \in \Omega_\alpha} \sum_{j \ne i} \sum_{k\ne i, k\ne j}\Big\{ \big[\bm f_{ij,ik}+\bm f_{ji,ik}\big]\cdot \bm v_i
 -\half \big[ \bm f_{ij,ik} +  \bm f_{ji,ik}\big] \cdot \bm v_i
-\half  \big[\bm f_{ij,jk} +  \bm f_{ji,jk} \big]\cdot \bm v_j \Big\}\\
 &= \half \sum_{i \in \Omega_\alpha} \sum_{j \ne i} \sum_{k\ne i, k\ne j}\Big\{ 
   \big[ \bm f_{ij,ik} +  \bm f_{ji,ik}\big] \cdot \bm v_i
 - \big[\bm f_{ij,jk} +  \bm f_{ji,jk} \big]\cdot \bm v_j \Big\}\\
&=\half \sum_{i \in \Omega_\alpha} \sum_{j \notin \Omega_\alpha} \sum_{k\ne i, k\ne j}\Big\{ 
  \big[ \bm f_{ij,ik} +  \bm f_{ji,ik}\big] \cdot \bm v_i
 -  \big[\bm f_{ij,jk} +  \bm f_{ji,jk} \big]\cdot \bm v_j \Big\}  
 \end{aligned}
 \]

Again, we have used the same trick to eliminate the terms for which $j \in \Omega_\alpha$. Let us define the flux,
\begin{equation}\label{eq: q-II}
 \bm q_{\alpha,\beta}^{II}=\half \sum_{i \in \Omega_\alpha} \sum_{j \in \Omega_\beta}
 \sum_{k\ne i, k\ne j}\Big\{ 
  \big[ \bm f_{ij,ik} +  \bm f_{ji,ik}\big] \cdot \bm v_i
 -  \big[\bm f_{ij,jk} +  \bm f_{ji,jk} \big]\cdot \bm v_j \Big\} 
 \end{equation}
The intuition is that these forces represent the interaction of the pair $(i,j)$ with a third atom in a neighboring cell. 

\smallskip

We now collect the remaining terms. In the equation \eqref{eq: kin}, we change the dummy index $\ell$ to $j$, and
the remaining terms are,
\[
\begin{aligned}
\half \sum_{i \in \Omega_\alpha} \sum_{j \ne i} \sum_{k  \ne i, k \ne j} &\Big\{\big[\bm f_{k j,ik}+\bm f_{j k,ik}
+\bm f_{kj,ij}+\bm f_{j k,ij}\big] \cdot \bm v_i\\
 &-\big[\bm f_{ij,ki} + \bm f_{ji,ki} + \bm f_{ij, kj}+\bm f_{ji, kj}\big]\cdot \bm v_k \Big\}.
\end{aligned}
\]

We exchange the order of summation over $j$ and $k$, 
\[
\begin{aligned}
\half \sum_{i \in \Omega_\alpha} \sum_{k \ne i} \sum_{j  \ne i, j \ne k} &\Big\{\big[\bm f_{k j,ik}+\bm f_{j k,ik}
+\bm f_{kj,ij}+\bm f_{j k,ij}\big] \cdot \bm v_i\\
 &-\big[\bm f_{ij,ki} + \bm f_{ji,ki} + \bm f_{ij, kj}+\bm f_{ji, kj}\big]\cdot \bm v_k \Big\}\\
=\half \sum_{i \in \Omega_\alpha} \sum_{k \notin \Omega_\alpha} \sum_{j  \ne i, j \ne k} &\Big\{\big[\bm f_{k j,ik}+\bm f_{j k,ik}
+\bm f_{kj,ij}+\bm f_{j k,ij}\big] \cdot \bm v_i\\
 &-\big[\bm f_{ij,ki} + \bm f_{ji,ki} + \bm f_{ij, kj}+\bm f_{ji, kj}\big]\cdot \bm v_k \Big\}.
\end{aligned}
\]
As a result, the flux only involves atoms near the boundary. To this end, we consider,
\begin{equation}\label{eq: q-III}
 \begin{aligned}
 \half \sum_{i \in \Omega_\alpha} \sum_{k \in \Omega_\beta} \sum_{j  \ne i, j \ne k} &\Big\{\big[\bm f_{k j,ik}+\bm f_{j k,ik}
+\bm f_{kj,ij}+\bm f_{j k,ij}\big] \cdot \bm v_i\\
 &-\big[\bm f_{ij,ki} + \bm f_{ji,ki} + \bm f_{ij, kj}+\bm f_{ji, kj}\big]\cdot \bm v_k \Big\}.
\end{aligned}
\end{equation}

To incorporate these formulas into the algorithm, we split the summation into,
\begin{equation}\label{eq: 4-terms}
\begin{aligned}
 &\half\sum_{i \in \Omega_\alpha} \sum_{k \in \Omega_\beta} \sum_{j \in \Omega_\alpha, j  \ne i}
\big[\bm f_{k j,ik}+\bm f_{j k,ik}+\bm f_{kj,ij}+\bm f_{j k,ij}\big] \cdot \bm v_i\\
& + \half\sum_{i \in \Omega_\alpha} \sum_{k \in \Omega_\beta} \sum_{j \in \Omega_\beta, j \ne k}
\big[\bm f_{k j,ik}+\bm f_{j k,ik}
+\bm f_{kj,ij}+\bm f_{j k,ij}\big] \cdot \bm v_i \\
& - \half\sum_{i \in \Omega_\alpha} \sum_{k \in \Omega_\beta} \sum_{j \in \Omega_\alpha, j  \ne i}
\big[\bm f_{ij,ki} + \bm f_{ji,ki} + \bm f_{ij, kj}+\bm f_{ji, kj}\big]\cdot \bm v_k\\
&-\half\sum_{i \in \Omega_\alpha} \sum_{k \in \Omega_\beta} \sum_{j \in \Omega_\beta, j \ne k}
\big[\bm f_{ij,ki} + \bm f_{ji,ki} + \bm f_{ij, kj}+\bm f_{ji, kj}\big]\cdot \bm v_k.
\end{aligned}
\end{equation}

 For the second and third terms, we define,
\begin{equation}\label{eq: q-III'}
\begin{aligned}
 \bm q_{\alpha,\beta}^{III}=&\half\sum_{i \in \Omega_\alpha} \sum_{k, j \in \Omega_\beta, j \ne k}
\big[\bm f_{k j,ik}+\bm f_{j k,ik}
+\bm f_{kj,ij}+\bm f_{j k,ij}\big] \cdot \bm v_i \\
& - \half\sum_{i, j \in \Omega_\alpha, j\ne i} \sum_{k \in \Omega_\beta} 
\big[\bm f_{ij,ki} + \bm f_{ji,ki} + \bm f_{ij, kj}+\bm f_{ji, kj}\big]\cdot \bm v_k.
\end{aligned}
\end{equation}
In addition, for the first sum, we can change:$j \to i$, $k \to j$ and $i \to k$, and combine it with the fourth term,
\begin{equation}\label{eq: q-IV}
\begin{aligned}
 \bm q_{\alpha,\beta}^{IV} =&  \half\sum_{i \in \Omega_\alpha} \sum_{j \in \Omega_\beta} \sum_{k \in \Omega_\alpha}  \big[\bm f_{ij,ki} + \bm f_{ji,ki} + \bm f_{ij, kj}+\bm f_{ji, kj}\big]\cdot \bm v_k \\
 & -\half\sum_{i \in \Omega_\alpha} \sum_{k \in \Omega_\beta} \sum_{j \in \Omega_\beta, j \ne k}
\big[\bm f_{ij,ki} + \bm f_{ji,ki} + \bm f_{ij, kj}+\bm f_{ji, kj}\big]\cdot \bm v_k.
 \end{aligned}
 \end{equation}

As the last step, we combine \eqref{eq: q-II} and \eqref{eq: q-IV},
\begin{equation}\label{eq: q-V}
 \begin{aligned}
  \bm q_{\alpha,\beta}^{II} +  \bm q_{\alpha,\beta}^{IV} =& 
   \half\sum_{i \in \Omega_\alpha} \sum_{j \in \Omega_\beta} \sum_{k \in \Omega_\alpha}  
   \big[\bm f_{ij,ik} + \bm f_{ji,ik}\big]\cdot\big(\bm v_i - \bm v_k)\\
&-   \half\sum_{i \in \Omega_\alpha} \sum_{j \in \Omega_\beta} \sum_{k \in \Omega_\alpha}  
         \big[\bm f_{ij, jk}+\bm f_{ji, jk} \big]\cdot\big(\bm v_j + \bm v_k) \\
    & + \half\sum_{i \in \Omega_\alpha} \sum_{j \in \Omega_\beta} \sum_{k \in \Omega_\beta}
            \Big\{ \big[\bm f_{ij,ik} + \bm f_{ji,ik}\big]\cdot\big(\bm v_i + \bm v_k)\\
 &        - \half\sum_{i \in \Omega_\alpha} \sum_{j \in \Omega_\beta} \sum_{k \in \Omega_\beta}
          \big[\bm f_{ij, jk}+\bm f_{ji, jk} \big]\cdot\big(\bm v_j - \bm v_k).
  \end{aligned}       
\end{equation}

The total flux is defined by combining the equation \eqref{eq: q-I}, \eqref{eq: q-III'} and \eqref{eq: q-V}. 
These formulas might appear to be complicated. But the implementation is quite straightforward. In the Appendix, we provide a pseudo-code for this algorithm. 

\subsection{Numerical Tests}\label{sec: tests}
In this section, we will show numerical tests to verify the consistency of our definitions of the traction and energy flux with the conservation laws for the Tersoff potential. 

In the computer experiments, we consider a system of silicon atoms with diamond structure. The lattice spacing is given by $a_0 = 5.43183 \AA$ at zero temperature.  The system contains $24000$ atoms with periodic boundary condition in all directions. The bulk is equally divided into $7$ blocks in the horizontal direction. Following a MD simulation, we calculate the total momentum and energy in each block, along with the traction and energy flux across the interfaces of the blocks. 

We choose a random initial velocity by setting the temperature of the system to be $30$ K. The Verlet's algorithm \cite{rapaport2004art} is used to generate the trajectory of the atoms. The time step $\Delta t$ is $5.395e10^{-5}$ fs, which is much smaller than the time step typically used in molecular dynamics so that the numerical error plays no role. We examine the consistency by checking the following equations,
\begin{equation}
  \label{eq:conservation}
  \begin{aligned}
    \bm t_{\alpha+\frac{1}{2}}(\tau) - \bm t_{\alpha-\frac{1}{2}}(\tau)&= \lim_{\Delta t \to 0} \frac{\mb p_{\alpha}(\tau+\frac{1}{2}\Delta t)-\mb p_{\alpha}(\tau-\frac{1}{2}\Delta t)}{\Delta t}\\
    q_{\alpha+\frac{1}{2}}(\tau)  - q_{\alpha-\frac{1}{2}}(\tau) &= \lim_{\Delta t \to 0} \frac{E_{\alpha}(\tau+\frac{1}{2}\Delta t) -E_{\alpha}(\tau-\frac{1}{2}\Delta t)}{\Delta t}
  \end{aligned}
\end{equation}
where $\bm t_{\alpha+\frac{1}{2}}= \bm t_{\alpha,\alpha+1}$ and $q_{\alpha+\frac{1}{2}}= q_{\alpha,\alpha+1}$  are the traction and energy flux between the $\alpha$-th and $\alpha+1$th cells. 

\begin{figure}[htbp]
    \begin{center}
  \includegraphics[width=0.45\textwidth]{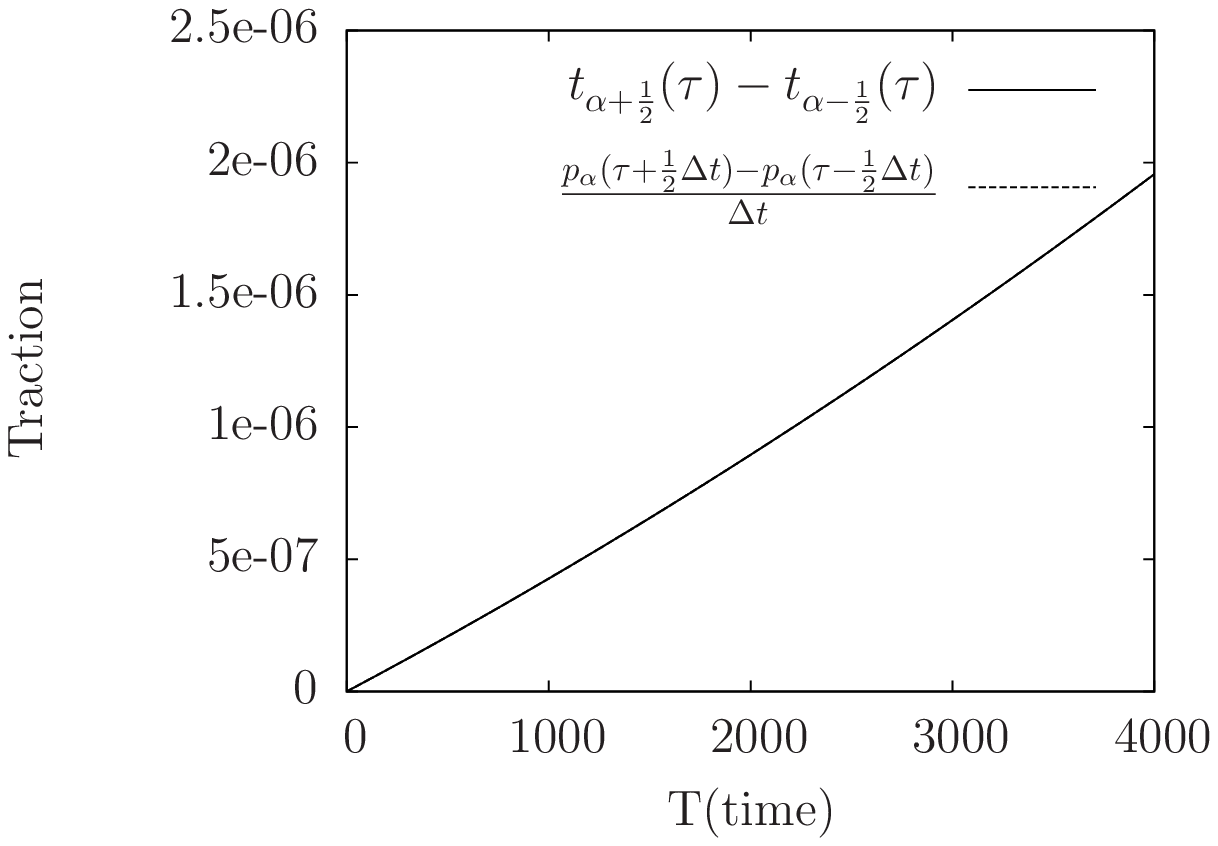}
  \includegraphics[width=0.45\textwidth]{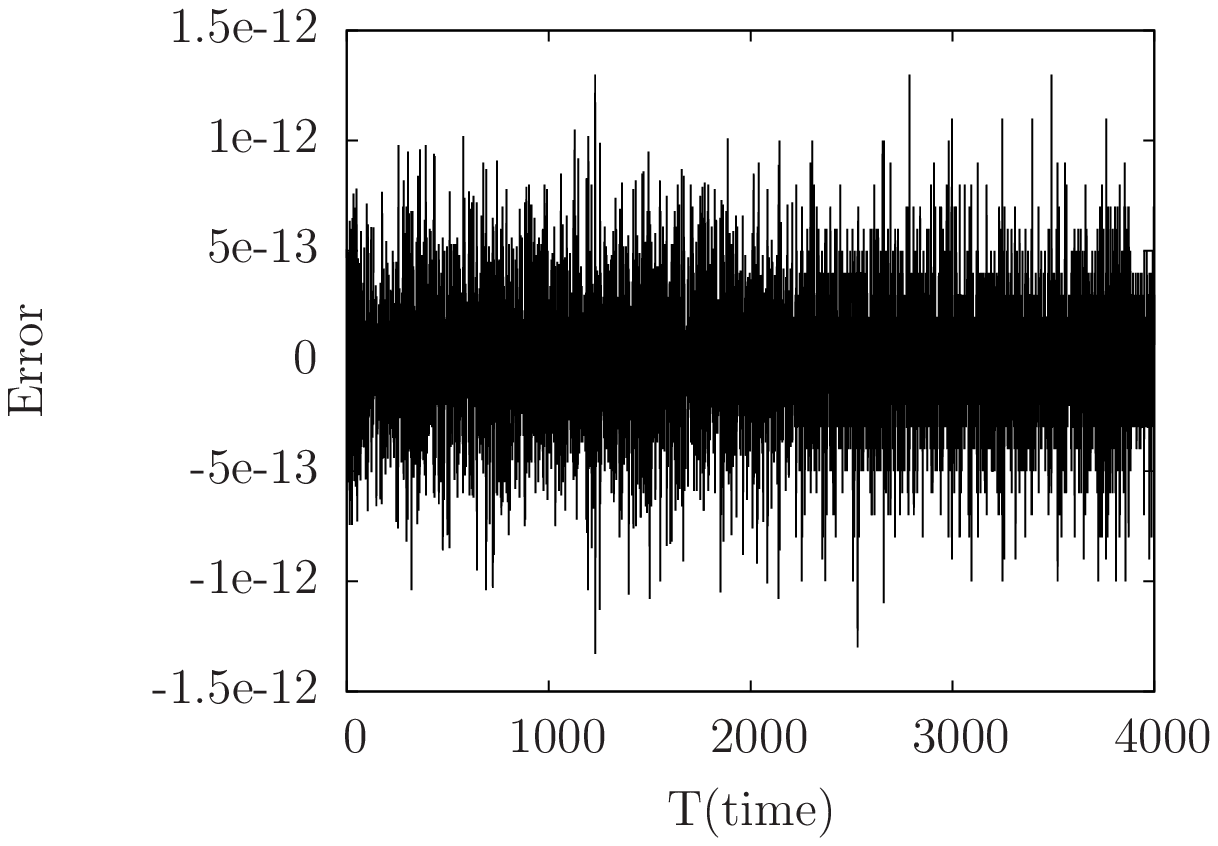}
  \includegraphics[width=0.45\textwidth]{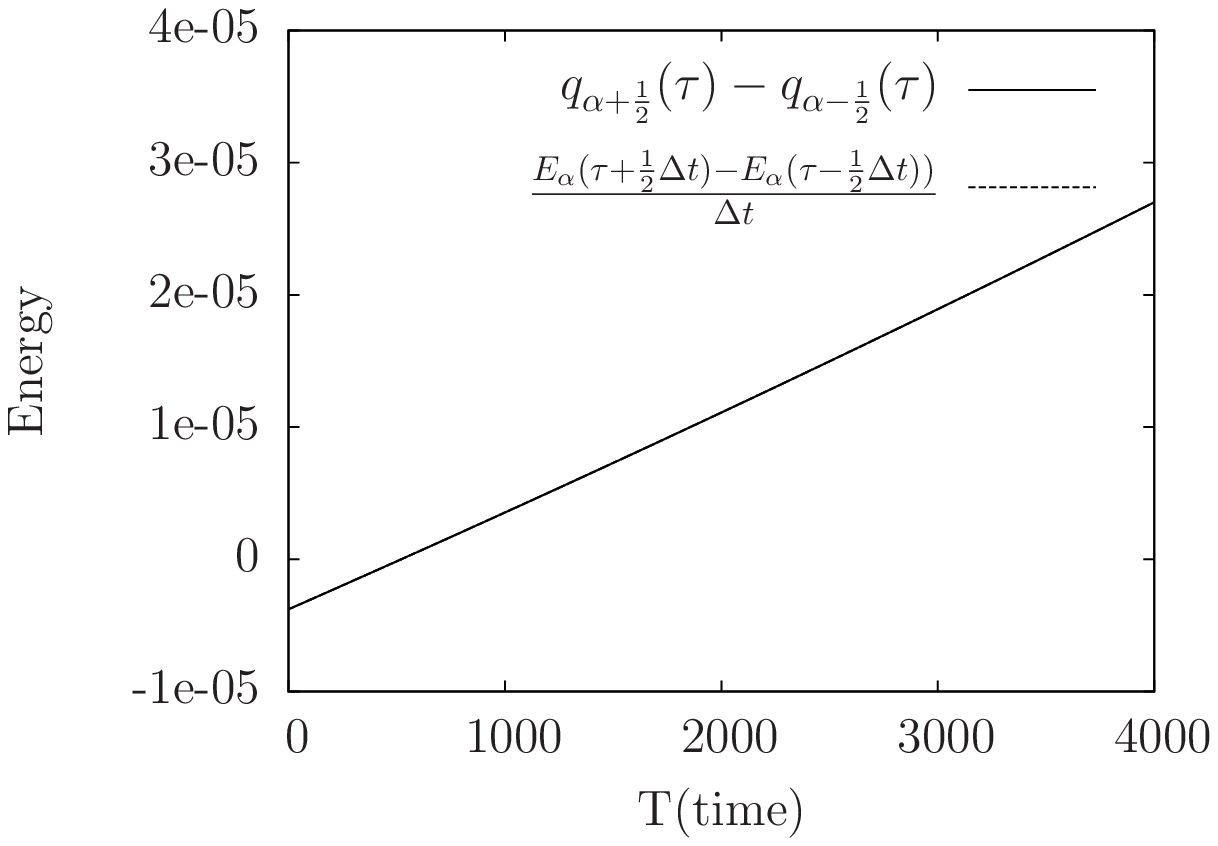}
  \includegraphics[width=0.45\textwidth]{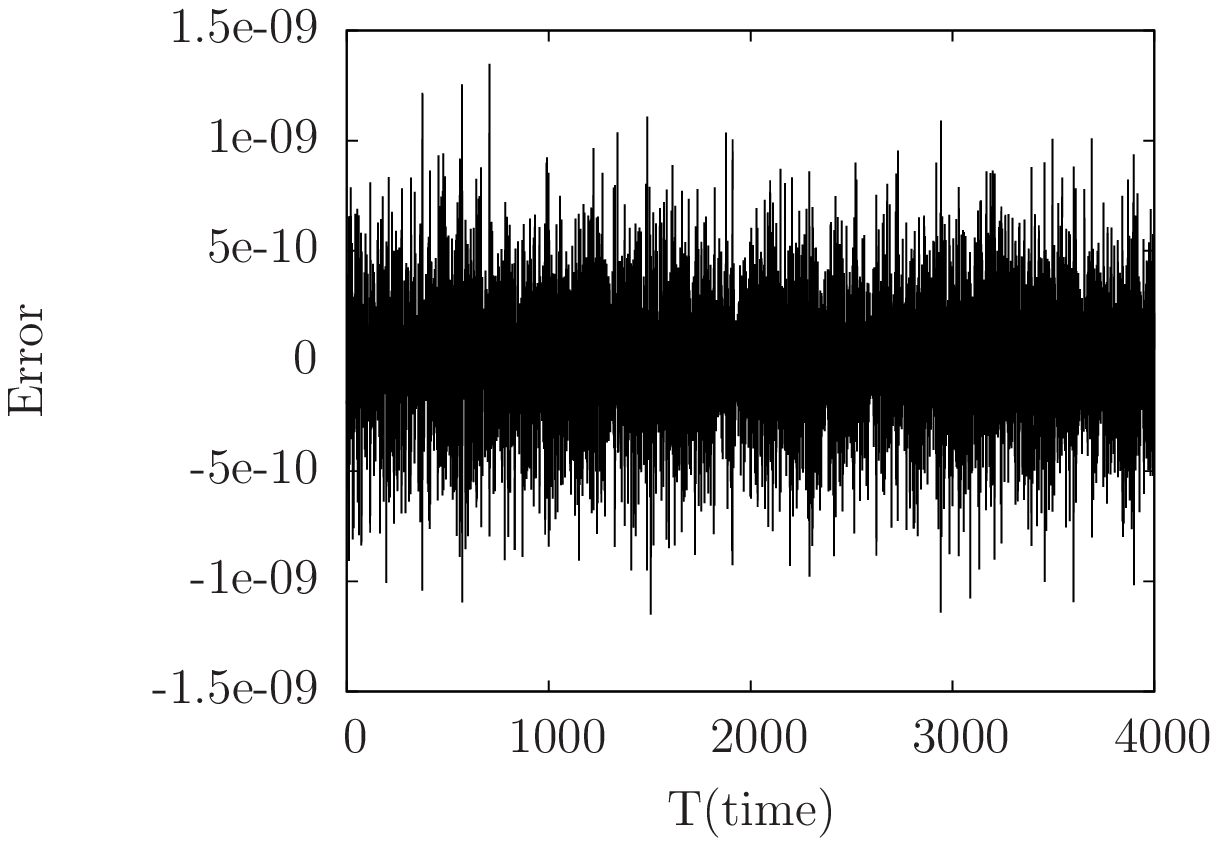}
  \end{center}
  \caption{Tests on momentum and energy conservation. Top left: the change of the momentum in a cell, compared to the fluxes in and out of the cell; Top right: the actual difference; Bottom left: the change of the energy in a cell, compared to the energy fluxes in and out of the cell; Bottom right: the actual difference. }
  \label{fig:energy_conservation}
\end{figure}

 The results, as summarized in Figure \ref{fig:energy_conservation}, demonstrate the consistency with \eqref{eq:conservation}. 

\bigskip

\section{The consistency with continuum mechanics models}\label{sec: consis2}

Having discussed the consistency with the conservation laws, we now show another criterion 
for the consistency. 
For a crystalline system with smooth displacement, it has been proved \cite{ArGr05,Weinan2007cauchy} that the corresponding
continuum limit is the elasticity model with constitutive relation given by the Cauchy-Born rule \cite{BlLeBrLi02,CB84,FrTh02}.  A simple description of the Cauchy-Born rule is as follows: Given the deformation gradient $A$, one can follow the uniform deformation gradient and define
an affine displacement field, from which the energy density \(W_\text{CB}(A)\) is defined. 
$W_\text{CB}(A)$ is the potential energy per unit volume. Further, the elastic stress is given by,
\begin{equation}
 P= \frac{\partial W_\text{CB}}{\partial A}.
\end{equation}
This is known as the Piola-Kirchhoff stress. For any interface, the total traction per unit area
would be given by \(\bm t_\text{CB}=P  \bm n\), with $\bm n$ being the unit normal.  An important observation is that this definition, in its original form, 
requires no force or energy decomposition. Therefore, it provides a unique reference to be compared to.  

\begin{figure}[!htb]
  \begin{center}
  \begin{subfigure}[b]{0.3\textwidth}
    \includegraphics[width=\textwidth]{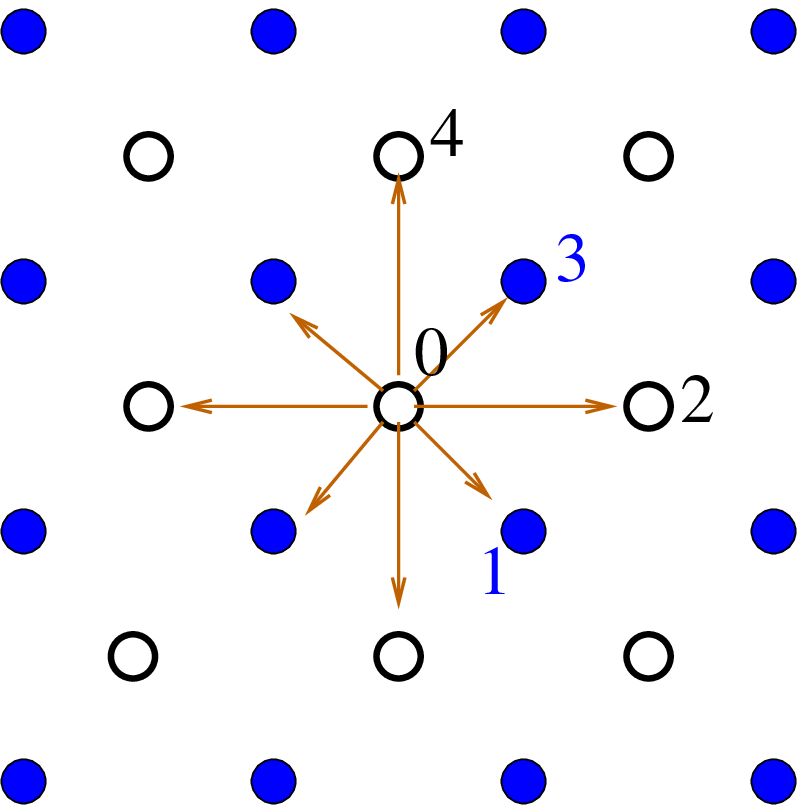}
  \end{subfigure}
  \qquad
  \qquad
  \begin{subfigure}[b]{0.3\textwidth}
    \includegraphics[width=\textwidth]{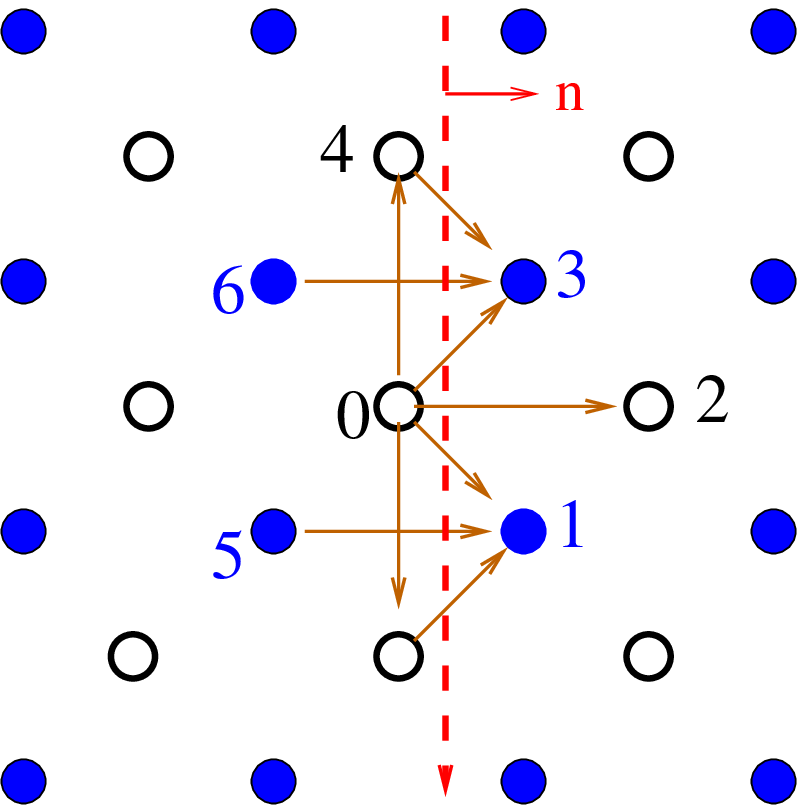}
  \end{subfigure}
  \end{center}
  \caption{A bcc lattice projected onto the (001) plane. Open circles are the atoms on the plane, and
  the filled circles indicate atoms above and below the plane with distance $a_0/2.$
  Left: The neighboring atoms
  for an atom in the bulk; Right: The atoms near an interface with normal $\bm n=(1,0,0)$.}
  \label{fig:cb2d}
\end{figure}

In order to demonstrate the comparison procedure with the Cauchy-Born elasticity, let us consider the $(100)$ plane in a bcc lattice. The projected atoms to the plane (001) is shown in Figure \ref{fig:cb2d}. To keep our demonstration brief, we assume the interaction is among nearest and second nearest neighbors.  We compare the traction (per unit area) from the Cauchy-Born model and with our definition for a system with uniform, but arbitrary deformation gradient. Fortunately, explicit expressions are available for the EAM model. 

Due to the uniform deformation gradient $A$, the traction corresponding to the Cauchy-Born rule can be written as
 \begin{equation}
 \label{eq:cb_stress}
    P = \frac{1}{\nu_0}\sum_i\bm f_{i0} \otimes \bm X_i,
\end{equation}
and
\begin{equation}
  \label{eq:cb_traction}
    \bm t_{CB} = \frac{1}{\nu_0}\sum_i\bm f_{i0} \bm X_i \cdot \bm n,
 \end{equation}
where $\nu_0$ is the volume per atom, and $\bm X_i$ represents the reference (undeformed) position of the $i$-th atom. 

Several comments need to be made to clarify this formula. First, the expression of the stress is derived from $P= \frac{\partial}{\partial A} W_\text{CB},$ and it is equivalent to the virial stress. 
Second, the force component $\bm f_{i0}$ comes from the force decomposition \eqref{eq: fij}. But in principle, this decomposition is not needed. In its original form, the elasticity energy density is defined as
\begin{equation}
 W_\text{CB}(A)= \lim_{|\Omega| \to +\infty} \frac{V(A\bm X_1, A\bm X_2, \cdots, A\bm X_N)}{|\Omega|},
\end{equation}
where $\Omega$ is the region occupied by the $N$ atoms. This part is clearly independent of the energy or force decomposition.  Upon taking the derivatives with respect to the deformation gradient, we have,
\[ \frac{\partial}{\partial A} V(A\bm X_1, A\bm X_2, \cdots, A\bm X_N)=
-\sum_i \bm f_i \otimes \bm X_i,\]
which, with any force decomposition \eqref{eq: fij}, becomes,
\[ \half \sum_i \sum_j \bm f_{ij} \otimes \bm X_{ij}.\]
This is where the force components can be introduced, and the formula is the better known expression of the Cauchy stress. At this point, any force decomposition would give the same result. 
But if we further assume that for the uniformly deformed state, the decomposed force $\bm f_{ij}$ only depends on the relative positions of the $i$th and $j$th atoms, then it suffices to consider 
the atoms around the zeroth atom, and simplify the expression to the form in \eqref{eq:cb_stress}. For the EAM potential, the force decomposition \eqref{eq: fu} does exhibit this translational invariance, since for the uniformly deformed state, the electron density is constant throughout the system. 

\smallskip

Now let's return to the bcc lattice, shown in Figure \ref{fig:cb2d}. The left panel illustrates
the neighboring atoms around the zeroth atom. Using the inversion symmetry, it is only necessary to consider the atoms with labels $0\sim 4$. Further, we used the open circles (0, 2, and 4) to indicate
the atoms on the plane, and the filled circles are the atoms above or below the planes with distance $a_0/2$. In the latter case, we label the atoms by 1, $\bar{1}$, 3,  $\bar{3}$ etc. From \eqref{eq:cb_stress}, the stress from the Cauchy-Born rule is given by,
\begin{equation}
  P= \frac{2}{a_0^3} \Big[\bm f_{0\bar{0}} \otimes \bm X_{0\bar{0}} +
  \bm f_{01}\otimes \bm X_{01}+
  \bm f_{0\bar{1}}\otimes \bm X_{0\bar{1}}+
  \bm f_{02}\otimes \bm X_{02}+
  \bm f_{03}\otimes \bm X_{03}+
  \bm f_{0\bar{3}}\otimes \bm X_{0\bar{3}}+
  \bm f_{04}\otimes \bm X_{04}\Big]. 
\end{equation}
The projection to the interface with normal vector $\bm n =(1,0,0) $ is given by,
\begin{equation}
\bm t_\text{CB}= \frac{1}{a_0^2} \Big[\bm f_{01} + \bm f_{0\bar{1}}
+ 2\bm f_{02} + \bm f_{03} + \bm f_{0\bar{3}}\Big].
\end{equation}

Meanwhile, the traction we defined in \eqref{eq: t-eam} can be calculated based on the right panel of 
Figure \ref{fig:cb2d}. More specifically, we have the traction per unit area,
\begin{equation}
  \label{eq:trac2d}
  \bm t = \frac{1}{a_0^2}(\bm f_{51}+ \mb f_{01} + \mb f_{0\bar{1}} + \bm f_{02} + \mb f_{01} + \mb f_{0\bar{1}} \Big].
\end{equation}
We chose these pairs since other pairs are replicas of these bonds. Again due to the translational symmetry, we have $\bm f_{51}=\bm f_{02},$ and the consistency of the two formulas are confirmed.  
It appears that this argument also applies to other potentials, and the only requirement is that the force decomposition satisfy translation symmetry for a uniformly deformed state.


\medskip

For complex lattices, the Cauchy-Born rule needs to be formulated with care. For clarity, we consider the case when there are two atoms in each primitive cell. In the Cauchy-Born rule,  the first atom in each unit cell would follow the uniform deformation gradient $A$, while the displacement of the second atom in each cell will be donated  by \(\bm p\) as an internal degree of freedom. The energy density in this case is written as $W(A,\bm p)$. The continuum elastic energy is then obtained 
by a minimization step,
\begin{equation}\label{eq:cbmodified}
  W_\text{CB}(A)=\min_{\bm p} W(A,\bm p).
\end{equation}

\begin{figure}[!htb]
  \centering
  \includegraphics[width = 0.45\textwidth]{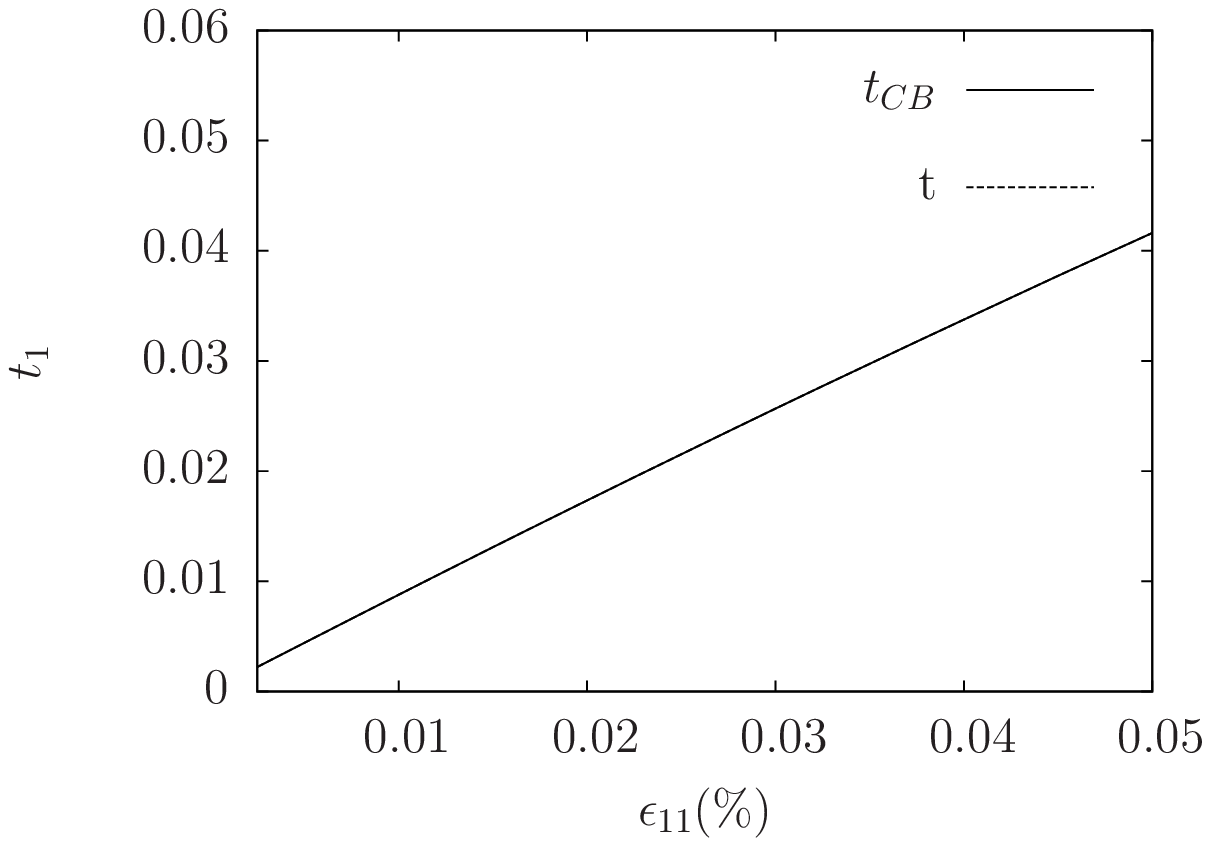}
  \includegraphics[width = 0.45\textwidth]{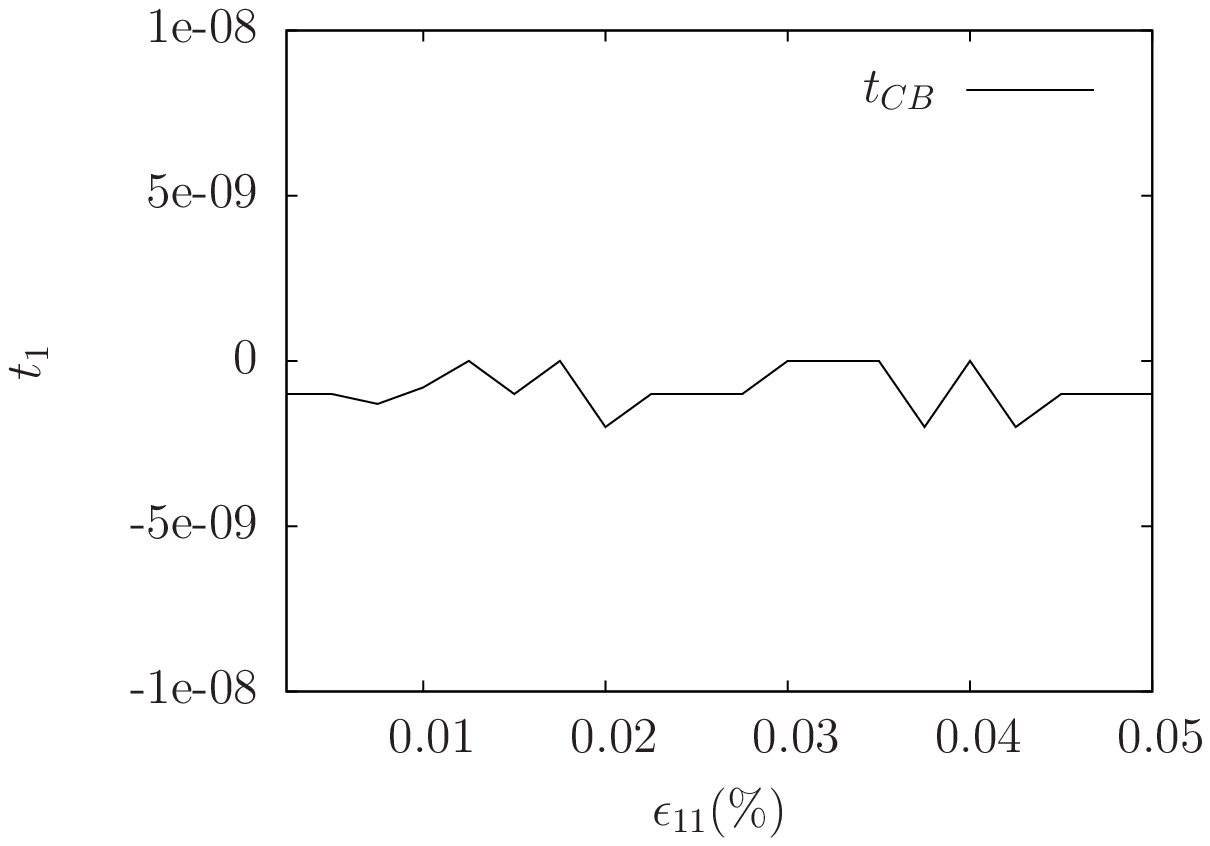}\\
  \includegraphics[width = 0.45\textwidth]{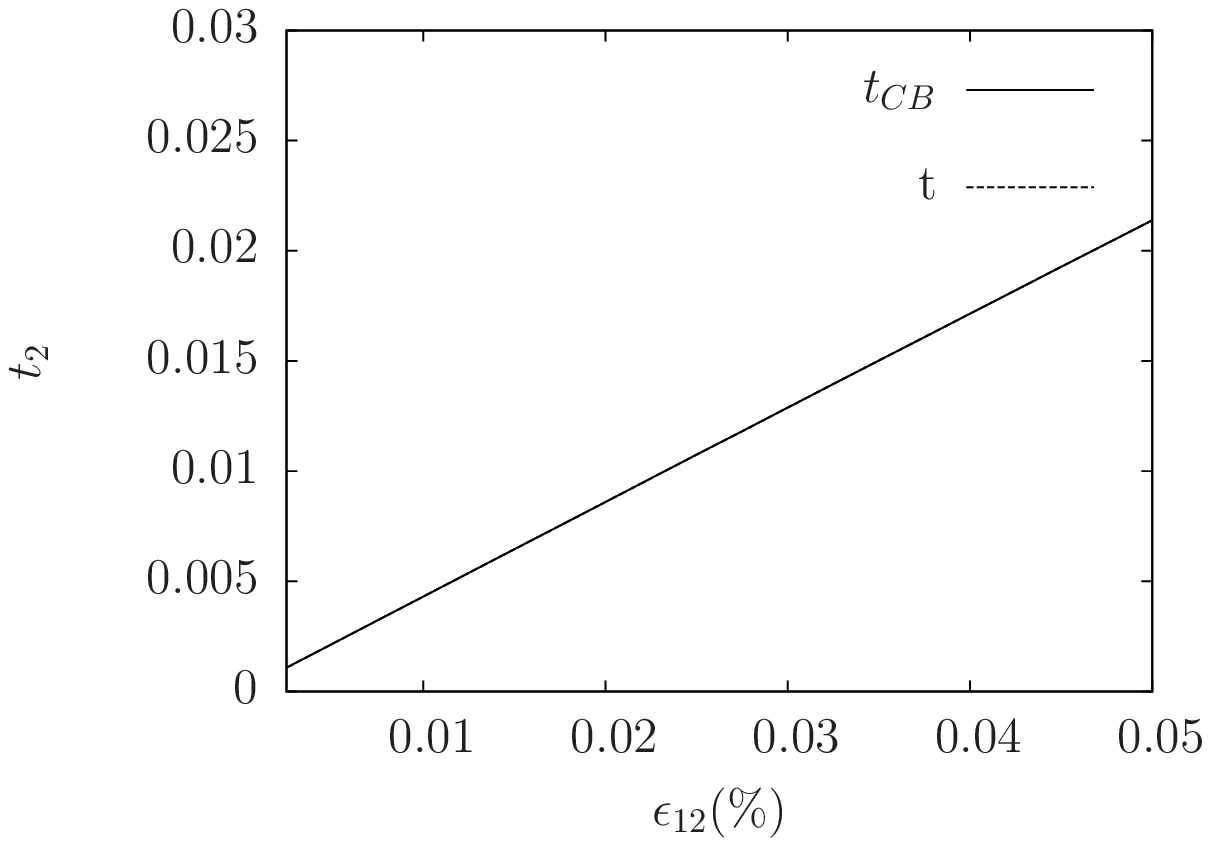}
  \includegraphics[width = 0.45\textwidth]{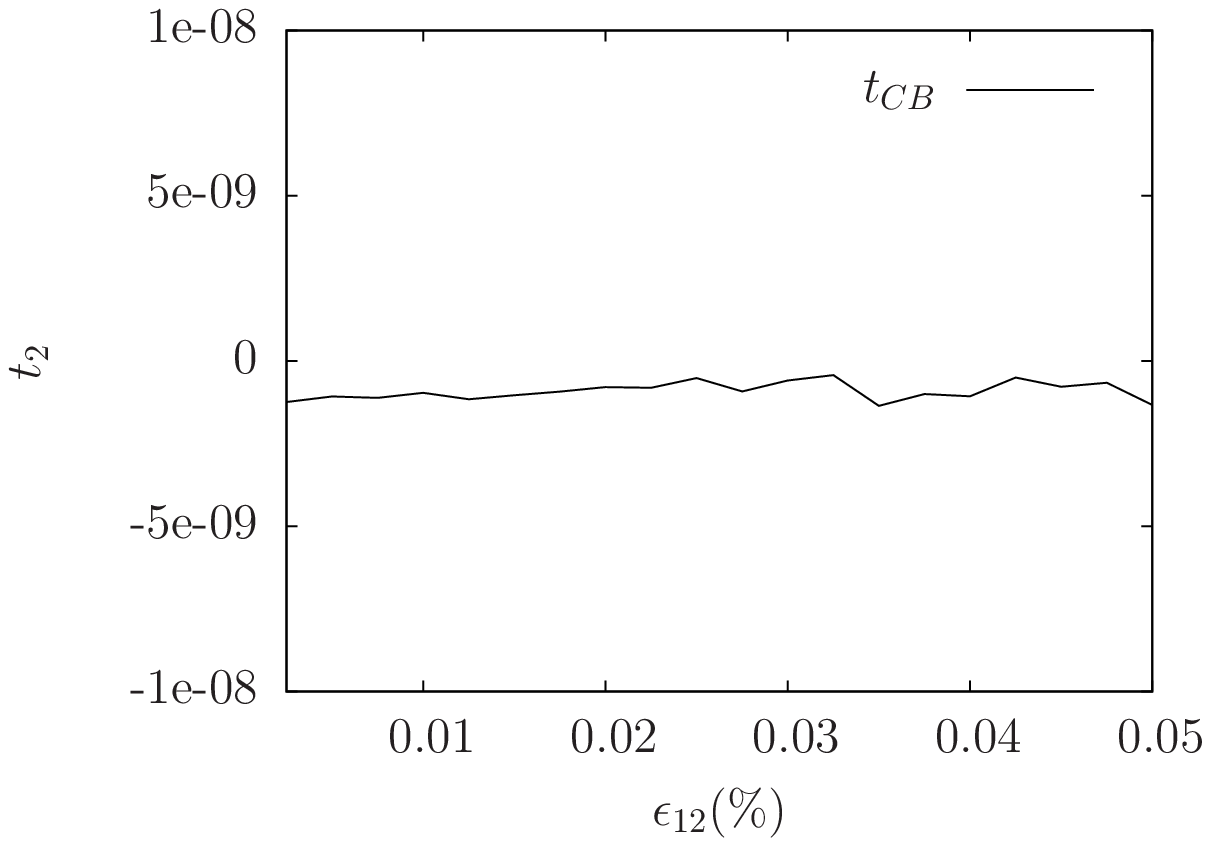}
  \caption{The traction determined from Cauchy-Born elasticity and our definition (Left panel). The actual difference (Right panel).}
  \label{cb_consistency}
\end{figure}

Now we consider the diamond structure of silicon modeled by the Tersoff potential. In this case, it would be a tedious procedure to verify the consistency by hand. Instead, we rely on a numerical test. For \eqref{eq:cbmodified}, we minimize $W(A,\bm p)$ using the BFGS method \cite{Liu1989lbfgs}.  Two tests, one with a uniform stretch (up to $5\%$), and the other with a uniform shear (up to $5\%$), are conducted, and the results are shown in Figure \ref{cb_consistency}. Clearly the agreement  is on the order of  machine precision.

To summarize briefly, we have shown that our definitions of the traction for both the EAM and the Tersoff potential are consistent with the continuum mechanics models. In general, our consistency criterion at this level is stated as follows: For any uniform deformation gradient $A$, and any rational planes, the traction per unit area is the same as $P \bm n$, with $P$ being the stress derived from the Cauchy-Born rule.

\begin{rem}
The current approach does not rule out other definitions of the traction or stress. In fact, in \cite{Admal2010unified}, an elegant idea was presented to construct other force decompositions. 
Our calculations based on Figure \ref{fig:cb2d} suggests an additional constraint on the force decomposition, at least for simple lattices. That is: For a uniformly deformed system, $\bm f_{ij}$
should exhibit translational invariance. For instance, the idea in \cite{Admal2010unified} relies on a fictitious term, which does not change the total energy, but gives different force decompositions. 
Suppose that this additional term is introduced uniformly along the interface, then the resulting traction would still be consistent. But if it is only introduced to some of the atoms, 
the traction may not be consistent according to our criterion.
\end{rem}

\begin{rem}
Thus far, we have only discussed the consistency of the traction. For the energy flux and in the absence of heat conduction, we expect that, as suggested by continuum mechanics models, the energy flux only contains the convection part: $q = \bm v^T P \bm n,$ where $\bm v$ is the macroscopic velocity.
In this case, our consistency criterion is stated as follows:  For a system with any uniform deformation gradient $A$ and any uniform velocity $\bm v$, the energy flux per unit area is the same as $\bm v^T P\bm n$, with $P$ being the stress derived from the Cauchy-Born rule. If the traction is consistent, then it is easy to see that the energy flux that we defined is also consistent in this sense  since the velocity can be factored out from the energy flux \eqref{eq: qeam}. The same holds for the Tersoff potential as well. A further comparison would be in the context of heat conduction. Such effort has been initiated in  the works  \cite{Schelling2002comparison,khadem2013comparison}. This issue, however, is generally very complicated and it will not be addressed in the current paper.   
\end{rem}

\section{A Nonhomogeneous example} 
\begin{figure}[htbp] 
   \includegraphics[width=0.5\textwidth]{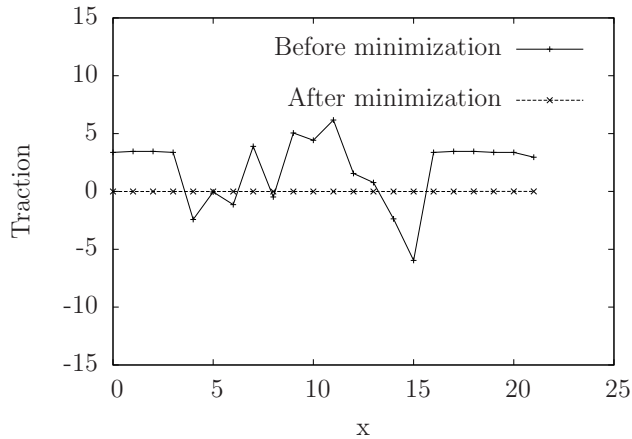} 
   \caption{A numerical test for a non-homogeneous system. The traction before and after minimization.}
   \label{fig:nonhomo}
\end{figure}

Unlike the homogenous case, the definition of the traction and energy flux for non-homogenous systems is difficult to validate. But physically, when a system is at a mechanical equilibrium, the traction should be zero all across the sample.  For the following example, we try to validate the definition of the traction in this particular setting: We consider a silicon system of dimension $20a_0\times20a_0\times20a_0$, with a void at the center. 
The radius of the void is $5a_0$.  The system is equally divided into $23$ blocks along the horizontal axis. Since each unit cell contains 8 atoms, the block size has been chosen to be less than $a_0$.  We initialize the system from a perfect diamond lattice, and create the void by removing the atoms in the middle. The atomistic model is the Tersoff potential \cite{Tersoff1989modeling}, and we  minimize the system to an equilibrium state by the BFGS method \cite{Liu1989lbfgs}. The atoms at the boundary are left free.  Figure \ref{fig:nonhomo} shows that the tractions initially are non-uniform along the $x$ axis. However, all the tractions settle to zero after the minimization steps. Therefore, they are consistent with the mechanical equilibrium state. 

\section{Summary and Discussion}

In this paper, we discussed the procedure for extracting average quantities based on the data from molecular dynamics simulations. In particular, we presented a control-volume representation of the fundamental conservation laws, from which the tractions and energy fluxes along the cell edges can be identified. Compared to formulations that are based on the continous form of the conservation laws \cite{Admal2010unified,Admal2011stress,Chen2006local,Hardy1982formulas,Webb2008reconsideration,Yang2014accurate,LiE05,Yang2012generalized,Zhou2005thermomechanical,Zhou2003new,Zimmerman2004calculation,Zimmerman2010material}, which usually lead to the divergence
of the stress and energy flux, the control-volume approach directly yields the traction and energy flux along a plane without the arbitrary additional divergence-free term. 

We chose to work with the EAM and Tersoff models because of their practical importance. The formulas that we have derived may not be new. Especially, it is unclear if they coincide with the formulas derived by Admal and Tadmor \cite{Admal2010unified,Admal2011stress}, since no explicit formulas were provided there for the Tersoff potential. Chen \cite{Chen2006local} already provided formulas based on the conservation of momentum and energy, but the Tersoff potential was mistreated as a three-body interaction. Our emphasis, however, is on whether these formulas are properly defined. This has become an issue, particularly when the non-uniqueness of the force decomposition was demonstrated in \cite{Admal2010unified,Admal2011stress}. In this paper, we proposed a two-level criteria to check the consistency
of these defined quantities. We first enforce that the conservation laws be exactly satisfied. This procedure has to be repeated if a new interatomic potential is considered, and one should not follow a naive generalization of the formulas derived from a different potentials. The second imposed condition is the consistency with the continuum model, with the constitutive relation given by the Cauchy-Born rule. To our knowledge, this  additional criterion has not been put forth in the literature. It is possible however that for every empirical potential, there is a natural force and energy decomposition, and the traction and energy fluxes defined based on such decomposition satisfy both of the criteria listed here. But at least, by verifying the two conditions, one can use these formulas with confidence.   

\bigskip

\section{Appendix}
\subsection{Pseudo code for computing the traction (Tersoff Potential)}
\begin{algorithmic}[1]
  \State $\bm t_{\alpha,\beta}=0$; 
  \State $\bm t_{\beta,\alpha}=0$; 
  \For{$i \ne j$}
  \If{$i \in \Omega_\alpha$ {\bf and} $j \in \Omega_\beta$ {\bf and} $\Omega_\alpha \cap \Omega_\beta \ne \emptyset$}
  \State \(\bm t_{\alpha,\beta} \leftarrow   \bm t_{\alpha,\beta} + \bm f_{ij,ij} \); 
  \State \(\bm t_{\beta,\alpha} \leftarrow   \bm t_{\beta,\alpha} - \bm f_{ij,ij}  \);  \Comment From 1st and 2nd terms eq \eqref{eq: t-I}
  \EndIf
  \For{$k=1,2,\cdots, N$} 
  \If{ $k\in \Omega_\alpha$ {\bf and} $k \ne i$}
  \State $\bm t_{\alpha,\beta} \leftarrow  \bm t_{\alpha,\beta} - \bm f_{ij, jk} $;
  \State $\bm t_{\beta,\alpha} \leftarrow  \bm t_{\beta,\alpha} + \bm f_{ij, jk} $;  \Comment From the 2nd sum in \eqref{eq: t-III}    
  \EndIf 
  \If{ $k\in \Omega_\beta$ {\bf and} $k \ne j$}
  \State $\bm t_{\alpha,\beta} \leftarrow  \bm t_{\alpha,\beta} +\bm f_{ij,ik} $;       
  \State $\bm t_{\beta,\alpha} \leftarrow  \bm t_{\beta,\alpha} -\bm f_{ij,ik}  $; \Comment From the 1st sum in \eqref{eq: t-III}     
  \EndIf
  \EndFor 
  \If{$i \in \Omega_\alpha$ {\bf and} $j \in \Omega_\alpha$}
  \For{$k=1,2,\cdots, N$} 
  \If{ $k\in \Omega_\beta$ {\bf and} $k \ne j$}
  \State $\bm t_{\alpha,\beta} \leftarrow  \bm t_{\alpha,\beta} + \bm f_{ij,ik}  + \bm f_{ij,jk} $;
  \State $\bm t_{\beta,\alpha} \leftarrow  \bm t_{\beta,\alpha} - \bm f_{ij,ik}  - \bm f_{ij,jk}   $;   \Comment From 1st and 2nd terms in eq \eqref{eq: t-II}     
  \EndIf
  \EndFor  
  \EndIf
  \EndFor
\end{algorithmic}

\subsection{Pseudo code for computing the energy flux (Tersoff Potential)}
\begin{algorithmic}[1]
 \State $\bm q_{\alpha,\beta}=0$; 
 \State $\bm q_{\beta,\alpha}=0$; 
 \For{$i \ne j$}
   \If{$i \in \Omega_\alpha$ {\bf and} $j \in \Omega_\beta$ {\bf and} $\Omega_\alpha \cap \Omega_\beta \ne \emptyset$}
   \State \(\bm q_{\alpha,\beta} \leftarrow   \bm q_{\alpha,\beta} + \bm f_{ij,ij}\cdot \big(\bm v_i + \cdot \bm v_j\big)  \); 
   \State \(\bm q_{\beta,\alpha} \leftarrow   \bm q_{\beta,\alpha} - \bm f_{ij,ij}\cdot \big(\bm v_i + \cdot \bm v_j\big)  \); \Comment From equations \eqref{eq: q-I}
   \EndIf
   \For{$k=1,2,\cdots, N$} 
   \If{ $k\in \Omega_\alpha$ {\bf and} $k \ne i$}
   \State $\bm q_{\alpha,\beta} \leftarrow  \bm q_{\alpha,\beta} + \bm f_{ij,ik} \cdot \big(\bm v_j- \bm v_k)- \bm f_{ij, jk} \cdot\big(\bm v_j + \bm v_k)$;
   \State $\bm q_{\beta,\alpha} \leftarrow  \bm q_{\beta,\alpha}  - \bm f_{ij,ik} \cdot \big(\bm v_j- \bm v_k)+\bm f_{ij, jk} \cdot\big(\bm v_j + \bm v_k)$; 
   \State \Comment From the 1st and 2nd terms in \eqref{eq: q-V}     
   \EndIf
   \If{ $k\in \Omega_\beta$ {\bf and} $k \ne j$}
   \State $\bm q_{\alpha,\beta} \leftarrow  \bm q_{\alpha,\beta} +\bm f_{ij,ik} \cdot\big(\bm v_i + \bm v_k) -\bm f_{ij,jk} \cdot\big(\bm v_j- \bm v_k) $;      
   \State $\bm q_{\beta,\alpha} \leftarrow  \bm q_{\beta,\alpha} -\bm f_{ij,ik} \cdot\big(\bm v_i + \bm v_k) +\bm f_{ij,jk} \cdot\big(\bm v_j- \bm v_k) $;    
   \State \Comment From the 3rd and 4th terms in \eqref{eq: q-V}
   \EndIf
   \EndFor
   \If{$i \in \Omega_\alpha$ {\bf and} $j \in \Omega_\alpha$}
   \For{$k=1,2,\cdots, N$} 
   \If{ $k\in \Omega_\beta$ {\bf and} $k \ne j$}
   \State $\bm q_{\alpha,\beta} \leftarrow  \bm q_{\alpha,\beta} +2\big[\bm f_{ij,ik}  + \bm f_{ij,jk}\big] \cdot \bm v_k $;
   \Comment From the 2nd term in \eqref{eq: q-III'}
   \State $\bm q_{\beta,\alpha} \leftarrow  \bm q_{\beta,\alpha} -2\big[\bm f_{ij,jk}  + \bm f_{ij,jk}\big] \cdot \bm v_k $;      
   \Comment From the 1st term in \eqref{eq: q-III'}
   \EndIf   
   \EndFor
   \EndIf
 \EndFor
 \State $\bm q_{\alpha,\beta} \leftarrow\bm q_{\alpha,\beta}/2$\;
\end{algorithmic}

\bibliographystyle{abbrv}
\bibliography{traction}

\end{document}